\documentclass[letter]{aa}  

\usepackage{graphicx}
\usepackage{txfonts}
\usepackage{lscape}
\usepackage[colorlinks = true,
            linkcolor = blue,
            urlcolor  = black,
            citecolor = cyan]{hyperref}

\newcommand{\hi}{H\,\textsc{i}}

\newcommand {\kms} {\,{\rm km\,s}^{-1}}

\newcommand {\um} {\,{\mu\rm m}}
\newcommand {\kpc} {\,{\rm kpc}}

\newcommand {\kmskpc} {\,{\rm km\,s}^{-1}\,{\rm \kpc}^{-1}}

\newcommand {\de}{^{\circ}}

\newcommand {\msun}{\,{\rm M}_\odot}

\newcommand{\Myr}{\,{\rm Myr}}

\newcommand{\Udisc}{\Upsilon_{\rm disc}}
\newcommand{\Ubulge}{\Upsilon_{\rm bulge}}
\newcommand{\Uphot}{\Upsilon_{\rm phot}}
\newcommand{\Udyn}{\Upsilon_{\rm dyn}}

\newcommand{\bagpipes}{\textsc{Bagpipes}}
\def\apjl{ApJ}%
\def\apjs{ApJS}%
\def\ao{Appl.~Opt.}%
\def\aap{A\&A}%
\defcitealias{Posti+19a}{P19}
\defcitealias{Lelli+16c}{L16}
\defcitealias{Marasco+23a}{M23}

\begin{document} 
\title{Photometric versus dynamical stellar masses and their impact on scaling relations in nearby disc galaxies}
\titlerunning{Stellar masses in nearby disc galaxies}
\authorrunning{A. Marasco et al.}

\author{
A. Marasco\inst{1},
S.\,M. Fall\inst{2},
E.\,M. Di Teodoro\inst{3},
P.\,E. Mancera Pi\~na\inst{4}
}
\institute{INAF – Padova Astronomical Observatory, Vicolo dell’Osservatorio 5, I-35122 Padova, Italy\\
\email{antonino.marasco@inaf.it}
\and
Department of Physics and Astronomy, Johns Hopkins University, 3400 N. Charles Street, Baltimore, MD 21218, USA
\and
Dipartimento di Fisica e Astronomia, Università degli Studi di Firenze, via G. Sansone 1, I-50019 Sesto Fiorentino, Firenze, Italy
\and
Leiden Observatory, Leiden University, P.O. Box 9513, 2300 RA Leiden, The Netherlands
}
\date{Received 28 January 2025 / Accepted 24 February 2025}

\abstract
{
The study of scaling relations of disc galaxies and their evolution across cosmic time requires accurate estimates of galaxy stellar masses, $M_\star$, over broad redshift ranges.
While photometric $M_\star$ estimates ($M_{\rm phot}$) based on spectral energy distribution (SED) modelling methods are employed routinely at high-$z$, it is unclear to what extent these are compatible with dynamical $M_\star$ estimates ($M_{\rm dyn}$), available for nearby galaxies.
Here, we compare newly determined, SED-model-based $M_{\rm phot}$ with previously obtained $M_{\rm dyn}$ inferred via rotation curve decomposition techniques in a sample of $\sim100$ nearby galaxies from the SPARC database.
We find that the two mass estimates show a systematic agreement at the $\sim12\%$ ($0.05$ dex) level and a $\sim55\%$ ($0.22$ dex) scatter across almost $5$ dex in $M_\star$.
Our $M_{\rm phot}$ estimates correspond to mass-to-light ratios in the $3.6\um$ band that increase gradually with $3.6\um$ luminosity, as a consequence of the earlier (later) assembly history of high-mass (low-mass) disc galaxies.
The choice of using either $M_{\rm dyn}$ or $M_{\rm phot}$ has only a marginal impact on the slope and zero-point of the Tully-Fisher and Fall relations: the observed orthogonal scatter in both relations is virtually the same for the two methods, and indistinguishable from that derived using a constant mass-to-light ratio in the $3.6\um$ band.
$M_\star$ estimates based on the assumption that discs are marginally stable lead to the largest scatter in the scaling relations.}

\maketitle
%

\section{Introduction}\label{s:introduction}
Stellar mass, $M_\star$, rotation speed, $v_{\rm rot}$, and specific angular momentum, $j_\star$, are among the most basic properties of galaxies. 
They are linked together by well-known scaling laws: the \citet{TullyFisher77} relation (TFR, $M_\star$ vs $v_{\rm rot}$) and the \citet{Fall83} relation (FR, $j_\star$ vs $M_\star$).
Characterising these laws at different redshifts allows us to trace the build-up of galactic mass and angular momentum across cosmic time, and thus to guide and constrain our understanding of galaxy formation, from simple analytical models \citep{FallEfstathiou80,MMW98} to the most complex hydrodynamical simulations \citep{Grand+17,ElBadry+18,Rodriguez-Gomez+22}. 
These scaling laws are well determined in the local Universe \citep[e.g.][]{Lelli+16a,Posti+18}, but are less robust at higher redshifts, with different studies reporting discrepant slopes, amplitudes, and dispersions \citep{DiTeodoro+16,Ubler+17,Harrison+17,Marasco+19a,EspejoSalcedo+25}. 

The scaling laws require accurate estimates of $M_\star$, which can be derived by two complementary approaches: photometric and dynamical (hereafter $M_{\rm phot}$ and $M_{\rm dyn}$).
For the former, photometry in several wavebands is matched to a stellar population synthesis (SPS) model with an adjustable mass and star formation history (SFH).
This approach is based on the fact that the light observed at different wavelengths is dominated by stellar generations of different ages, and hence different mass-to-light ratios.
It has the advantage of requiring only broad-band imaging, and it is the only reliable method of estimating $M_\star$ at high redshifts.
Its main disadvantage is that it is subject to possible systematic uncertainties in the stellar initial mass function (IMF), stellar spectral libraries, and modelling of the advanced stages of stellar evolution \citep[e.g.][]{MartinezGarcia+21}.

The dynamical approach to estimating $M_\star$ is based on the motions of stars and gas in galaxies induced by gravity.
For disc galaxies, this usually entails fitting the observed rotation curve (nearly circular motion) to a mass model consisting of several components: a stellar disc and stellar bulge, a gaseous disc, and a dark matter (DM) halo.
Under ideal conditions, the masses of all four components can be uniquely determined. 
In practice, however, this is often not possible as a result of degeneracies among the masses, especially those of the flat components (stellar and gaseous discs) and the round components (stellar bulge and DM halo).
A great deal of experience in such mass modelling over the years has shown that flat-round (or `disc-halo') degeneracies are strong when the rotation curve covers only the stellar disc and are substantially less severe when it extends far beyond \citep[e.g.][]{vanAlbada+85}.
This in turn means that H$\alpha$ rotation curves alone are insufficient to break the degeneracy and that extended \hi\ or a combination of H$\alpha$ and \hi\ rotation curves are necessary, but not always sufficient \citep[e.g.][]{Kent86,Kent+87}.

The purpose of this Letter is to make a direct comparison between both estimates of stellar mass, $M_{\rm phot}$ and $M_{\rm dyn}$, for a local sample of $\sim100$ well-studied disc galaxies with extended \hi\ rotation curves.
In this way, we can assess any potential biases in estimates of $M_\star$ by the two approaches and their impact on scaling relations.
This comparison is only possible for present-day galaxies, but is crucial for evolutionary studies, which must rely on photometry alone to estimate $M_\star$ at higher redshifts.
Until now, this topic has received surprisingly little attention \citep[but see][for a possible exception]{Ponomareva+18}.
The ideal sample of galaxies for this study is the \emph{Spitzer} Photometry and Accurate Rotation Curves (SPARC) database \citep[][hereafter \citetalias{Lelli+16c}]{Lelli+16c}.
Galaxies in this sample already have well-determined estimates of $M_{\rm dyn}$ from \citet[][, hereafter \citetalias{Posti+19a}]{Posti+19a} and images in multiple wavebands, from which we derive new homogeneous estimates of $M_{\rm phot}$ in this paper.

\section{Stellar masses of the SPARC sample}\label{s:mstar}
The SPARC database includes $175$ nearby disc galaxies spanning a broad range of morphologies (from S0 to Irr) and $3.6\um$ luminosity, $L_{3.6}$ ($\sim5$ dex).
All galaxies have extended, high-quality rotation curves from previous \hi/H$\alpha$ studies and broad-band photometry in a variety of filters.
SPARC has already been used to explore scaling relations for disc galaxies in the local Universe such as the TFR \citep{Lelli+16a,Lelli+19} and the FR \citep{Posti+18, ManceraPina+21a,ManceraPina+21b}.
In this section, we describe our $M_{\rm dyn}$ and $M_{\rm phot}$ estimates for the SPARC sample, and outline two additional approaches that can be used to infer $M_\star$ in the absence of detailed kinematic modelling or multi-wavelength photometry.

    \subsection{Dynamical stellar masses}\label{ss:mstar_dyn}
The \citetalias{Posti+19a} estimates of $M_{\rm dyn}$ adopted here were derived by fitting a four-component mass model (with stellar disc and bulge, gaseous disc, and DM halo) to the observed H$\alpha$/\hi\ rotation curves of $110$ SPARC galaxies\footnote{In \citetalias{Posti+19a}, systems with poor rotation curve fit or with multi-modal posterior on the halo mass were discarded from the full SPARC sample, reducing the final number of galaxies from $175$ to $110$.}.
The observed \hi\ and $3.6\um$ surface brightness profiles were used to constrain the gas and stellar contributions, respectively, with the mass-to-light ratio of the stellar disc in the $3.6\um$ band ($\Udisc$) as the only free parameter, while bulges were assumed to have $\Ubulge\!=\!1.4\Udisc$, as was suggested by SPS models \citep[e.g.][]{SchombertMcGaugh14}.
The DM halos follow a \citet*{NFW} density profile, described by two parameters: virial mass ($M_{200}$) and concentration ($c_{200}$).
The free parameters of the model ($\Udisc$, $M_{200}$, $c_{200}$) were fitted to the data via Bayesian Markov chain Monte Carlo (MCMC) technique assuming a Gaussian prior on the $c_{200}$-$M_{200}$ relation derived from cosmological N-body simulations in a $\Lambda$CDM framework \citep{Dutton&Maccio14} and, most importantly, a flat prior on $\Udisc$ in the range $0.01-1.2$, which broadly encompasses the values inferred from SPS models.
We refer the reader to \citetalias{Posti+19a} for a more in-depth discussion of this method. 
The Bayesian approach inherently accounts for the disc-halo degeneracy, inflating the uncertainty on $\Udisc$ when the contribution of the stars to the rotation curve cannot be unambiguously separated from that of the DM. This is often the case for low-mass galaxies.

The $\Udisc$ obtained with this technique (listed in Table A.1. of \citetalias{Posti+19a}) are valid for the original $L_{3.6}$ of the SPARC galaxies reported by \citetalias{Lelli+16c}.
For consistency with our new photometry (see Section \ref{ss:mstar_photo}), we corrected $\Udisc$ by the ratio of the original-to-new $L_{3.6}$, so that the final mass, which is what is actually fitted to the rotation curve, is unchanged.
We finally computed $M_{\rm dyn}$ as
\begin{equation}\label{eq:Ms_from_MLratio}
    M_{\rm dyn} = \Udisc L_{3.6}(1-f_{\rm b}) + \Ubulge L_{3.6}f_{\rm b} = \Udisc L_{3.6}(1+0.4f_{\rm b})
,\end{equation}
where $f_{\rm b}$ is the bulge luminosity fraction of each system, determined by \citetalias{Lelli+16c} via disc-bulge decomposition of the $3.6\um$ light profile.
Equation (\ref{eq:Ms_from_MLratio}) can be written as $M_{\rm dyn}\!=\!\Udyn L_{3.6}$, where $\Udyn\equiv\Udisc(1+0.4f_{\rm b})$ is the total dynamical stellar mass-to-light ratio in the $3.6\um$ band.
Only $\sim10\%$ of the galaxies in SPARC have $f_{\rm b}\!>\!0.2$; hence, $\Udyn\!\simeq\!\Udisc$ in most cases.

The \citetalias{Posti+19a} estimates of $M_{\rm dyn}$ adopted here are similar to the ones derived for SPARC galaxies by \citet{Katz+17} and \citet{Li+20,Li+22}.
We prefer the \citetalias{Posti+19a} estimates because they lead to a monotonically rising stellar-to-halo mass relation for disc-dominated galaxies, as is required mathematically by the observed power law TFR and FR \citep{Posti+19b,PostiFall21,DiTeodoro+23}.

    \subsection{Photometric stellar masses}\label{ss:mstar_photo}
To estimate $M_{\rm phot}$ for the SPARC galaxies, we analysed publicly available archival images in multiple broad bands: near- and far-ultraviolet ($NUV$ and $FUV$) from the Galaxy Evolution Explorer \citep[\emph{GALEX;}][]{GildePaz+07}; \emph{g}, \emph{r}, and \emph{z} from the DESI Legacy Imaging Surveys \citep{Dey+19}; $J$, $H$, and $K_{\rm s}$ from the Two Micron All Sky Survey \citep[2MASS;][]{Skrutskie+06}; $3.6\um$ from the IRAC camera of the \emph{Spitzer Space Telescope} \citep{Werner+04}; and $22\um$ from the Wide-field Infrared Survey Explorer \citep[WISE;][]{Wright+10}. 
Our analysis of these images followed the approach described by \citet{Marasco+23a} and detailed here in Appendix \ref{a:procedure}.
This uniform reanalysis of all bands allowed us to build an homogeneous photometry database suitable for spectral energy distribution (SED) modelling. There is excellent agreement between the original $L_{3.6}$ derived by \citetalias{Lelli+16c} and our new measurements, as is shown in Fig.\,\ref{fig:L36_comparison}.
Our photometry catalogue is available at the CDS (Table C.1).

We inferred the SFHs of the galaxies by SED modelling.
This approach does not require decomposing the images into disc and bulge components to account correctly for their different SFHs, since all their light and mass are simply added together.
The SED modelling was done via the Bayesian Analysis of Galaxies for Physical Inference and Parameter EStimation (\bagpipes) code \citep{Carnall+18}, using only data that satisfy a series of quality flags and adopting a `non-parametric' SFH, whereby the SFR of a galaxy is approximated by a series of discrete step functions in each of the age bins.\footnote{The term `non-parametric' is actually a misnomer here because each of the discrete SFRs is a fitted parameter of the model.}
\bagpipes\ uses the \citet{BruzualCharlot03} SPS models (2016 version) with the \citet{KroupaBoily02} IMF. 
We adopted the dust attenuation model of \citet{CF00}.
The details of our SED modelling approach are provided in Appendix \ref{a:SED_modelling}.
Throughout this Letter, we quantify the match between the observed and best-fit model SED using a $\chi^2$ statistic, whereby we test the hypothesis that the data are a stochastic realisation of the model.
In practice, we computed $p(>\!\chi^2)$, the p value of the $\chi^2$ statistics with $n$ degrees of freedom, where $n$ is the number of bands used in the fit.
Low $p(>\!\chi^2)$ values indicate a mismatch between the model and the data that is unlikely to arise from photometric uncertainties alone.
This approach is suited to our study, as the number of SED bands varies from one galaxy to another and is often less than the number of free parameters used in the modelling.

With this method, we derived $M_{\rm phot}$ for $146$ SPARC galaxies directly from the output SFH of the best-fit model.
These estimates include stellar remnants but exclude the gas lost by stellar winds and supernovae in order to make direct comparisons with $M_{\rm dyn}$.
We define the corresponding photometric mass-to-light ratio in the $3.6\um$ band as $\Uphot\equiv M_{\rm phot}/L_{3.6}$.

    \subsection{Other methods}\label{ss:other_methods}
We considered two additional, simpler approaches to determining $M_\star$ which can be taken as addenda to the two methods discussed above.
The first is the `marginal stability condition' (MSC), which is based on the theoretical prediction that strong bar modes in stellar and gaseous discs are suppressed for $v^2_{\rm rot}\gtrsim G M_{\rm d}/R_{\rm d}$ \citep{Efstathiou+82, Christodoulou+95}, where $M_{\rm d}$ and $R_{\rm d}$ are the disc mass and exponential scale radius.\footnote{For a critical discussion of the MSC see \citet{Romeo+23}.}
Assuming that discs are marginally stable \citep[e.g.][]{Fall83,MMW98} leads to $M_{\rm d}\!\simeq\!M_\star+1.33 M_{\rm HI}\!\simeq\!v^2_{\rm rot} R_{\rm d}/G$, where the $1.33$ factor accounts for the He content in the ISM.
This formula is only approximate, since molecules are neglected and the stellar and gaseous discs are assumed to have the same radial profiles, but it can be used to infer $M_\star$ in our sample as all the  quantities required are readily available in the SPARC database.\footnote{We set $v_{\rm rot}\!=\!v_{\rm flat}$, the velocity in the flat part of the rotation curve.}

The second approach is based on the indication that the scatter in the TFR is minimised by assuming a constant $\Udisc$ in the $3.6\um$ band \citep[e.g.][]{Lelli+16a}.
The SPS models \citep[e.g.][]{McGaughSchombert14} suggest that the optimal value to use for $z\sim0$ disc galaxies is $\Udisc\!\simeq\!0.5$.
This approach has been used in previous studies \citep[e.g.][]{Posti+18,Lelli+19}.
The four $M_\star$ estimates, together with the other physical parameters used to build the TFR and FR for the galaxies studied here, are reported in Table C.2 (available at the CDS).

\begin{figure*}
\begin{center}
\includegraphics[width=0.85\textwidth]{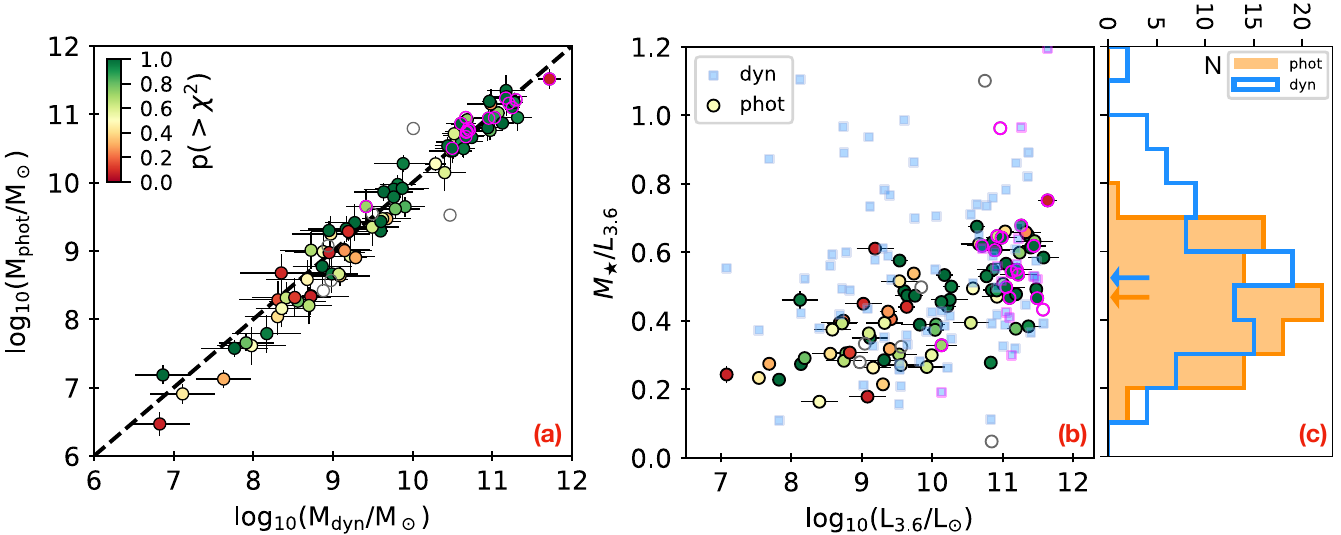}
\caption{($a$)--$M_{\rm phot}$ vs $M_{\rm dyn}$ for $95$ galaxies in the SPARC sample, colour-coded by $p(>\!\chi^2)$. Empty circles show galaxies with poor SED fits: $p(>\!\chi^2)\!<\!0.05$.
($b$)--$\Udyn$ (light blue squares) and $\Uphot$ (filled circles, colour-coded as in panel $a$) vs $L_{3.6}$. 
Error bars on $\Udyn$, not displayed for clarity, vary from $\pm0.3$ for the faintest galaxies to $\pm0.1$ for the brightest. $\Uphot$ increases gradually from $\sim0.2$ to $\sim0.7$ across more than $4$ dex in $L_{3.6}$. ($c$)--Distribution of $\Uphot$ (orange-filled histogram) and $\Udyn$ (blue empty histogram). Galaxies with poor SED fits have been removed. The arrows show the medians of the distributions. The purple-edged circles in panels $a$ and $b$ show systems with $f_{\rm b}>0.2$.}
\label{f:phot_vs_dyn}
\end{center}
\end{figure*}

\section{Results}\label{s:results}
\begin{figure*}
\begin{center}
\includegraphics[width=0.85\textwidth]{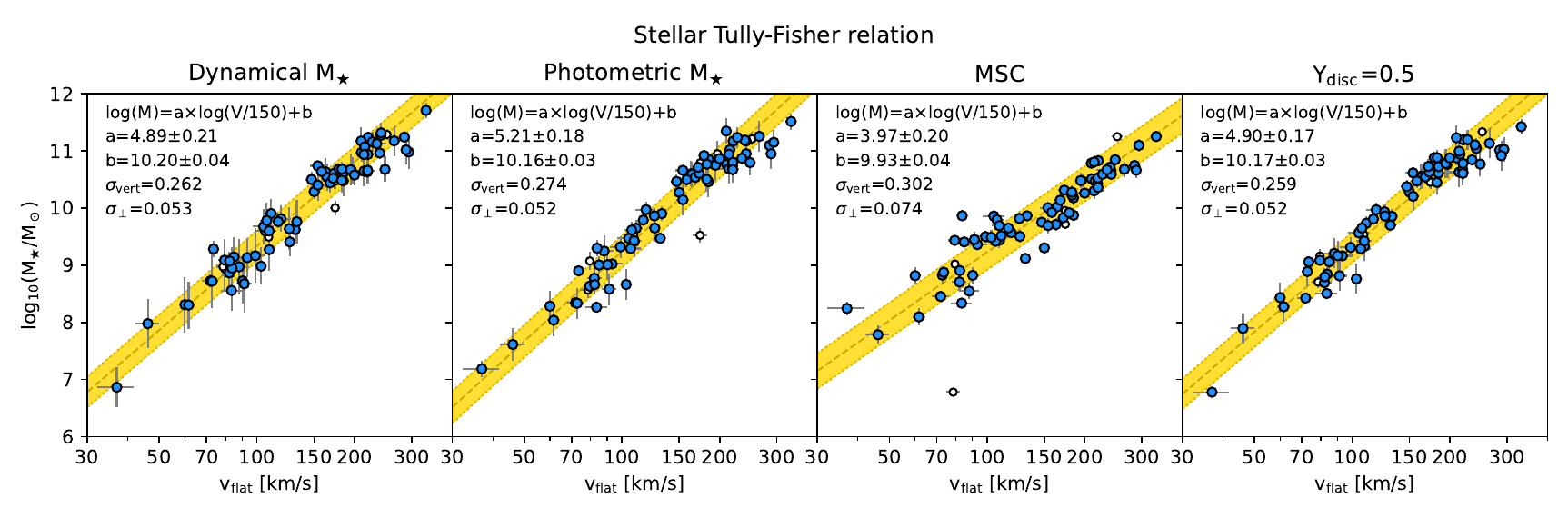}
\includegraphics[width=0.85\textwidth]{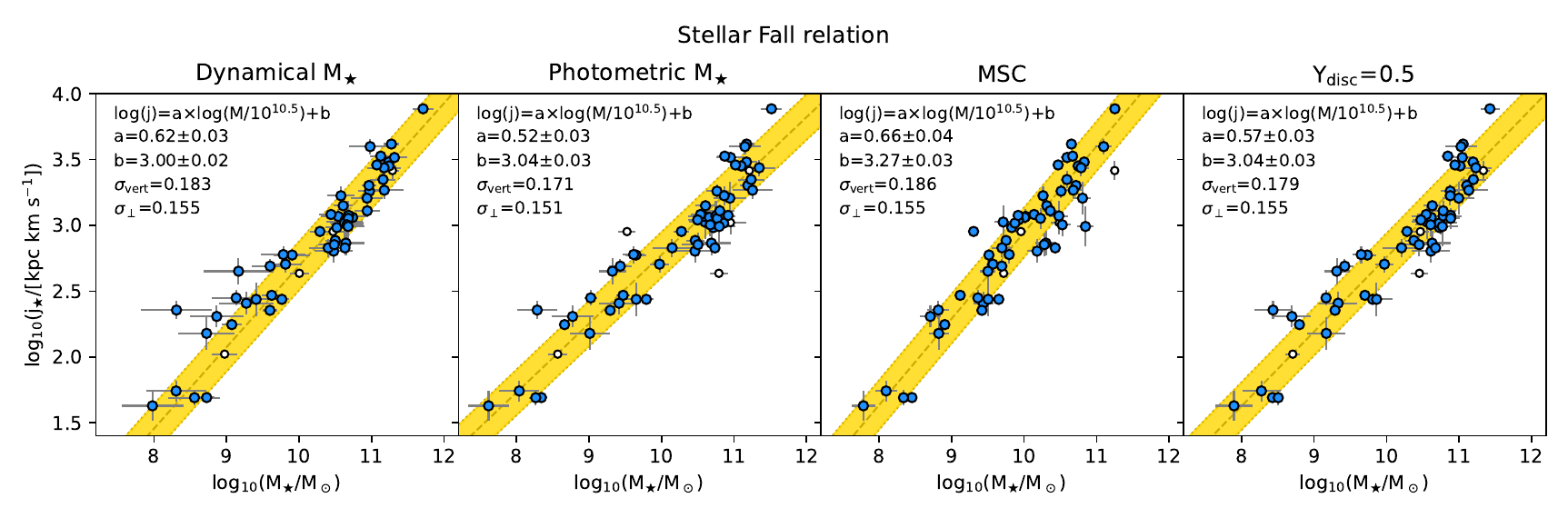}
\caption{Stellar Tully-Fisher relation (top panels) and Fall relation (bottom panels) for SPARC galaxies with $M_\star$ determined using the dynamical method (first column), the photometric method (second column), the MSC method (third column), and Equation (\ref{eq:Ms_from_MLratio}) with $\Udisc\!=\!0.5$ (fourth column). 
Only galaxies for which all four $M_\star$ estimates are available are shown.
Galaxies with $p(>\!\chi^2)$ above and below $0.05$ are shown with filled and open circles, respectively.
In each panel, the dashed black line shows the best linear fit to the filled circles, with the yellow band representing the vertical observed scatter ($\sigma_{M_\star}$ or $\sigma_{j_\star}$).
The best-fit parameters are reported in Table \ref{t:scaling_relations_params} and on the top left portion of each panel.}
\label{f:scaling_relations}
\end{center}
\end{figure*}

\begin{table}
\caption{Best-fit parameters for the stellar TFR and the stellar FR in the SPARC sample using different methods of estimating $M_\star$.}
\label{t:scaling_relations_params}
\centering
\begin{tabular}{lcccc}
\hline\hline
\noalign{\smallskip}
\multicolumn{5}{c}{TFR (68 galaxies), $x\!=\!v_{\rm flat}/150\kms$, $y\!=\!M_\star/\msun$} \\
method & $a$ & $b$ & $\sigma_{M_\star}$ & $\sigma_\perp$\\    
\hline
\emph{dyn}  &  $4.89\pm0.21$  &  $10.20\pm0.04$  &  $0.262$  &  $0.053$ \\
\emph{phot}  &  $5.21\pm0.18$  &  $10.16\pm0.03$  &  $0.274$  &  $0.052$ \\
\emph{MSC}  &  $3.97\pm0.20$  &  $9.93\pm0.04$  &  $0.302$  &  $0.074$ \\
$\Upsilon_{\rm disk}\!=\!0.5$  &  $4.90\pm0.17$  &  $10.17\pm0.03$  &  $0.259$  &  $0.052$\\
\hline\hline
\noalign{\smallskip}
\multicolumn{5}{c}{FR (55 galaxies), $x\!=\!M_\star/10^{10.5}\msun$, $y\!=\!j_\star/\kmskpc$}\\ 
method & $a$ & $b$ & $\sigma_{j_\star}$ & $\sigma_\perp$\\  
\hline
\emph{dyn}  &  $0.62\pm0.03$  &  $3.00\pm0.02$  &  $0.183$  &  $0.155$ \\
\emph{phot}  &  $0.52\pm0.03$  &  $3.04\pm0.03$  &  $0.171$  &  $0.151$ \\
\emph{MSC}  &  $0.66\pm0.04$  &  $3.27\pm0.03$  &  $0.186$  &  $0.155$ \\
$\Upsilon_{\rm disk}\!=\!0.5$  &  $0.57\pm0.03$  &  $3.04\pm0.03$  &  $0.179$  &  $0.155$ \\
\hline\hline
\noalign{\smallskip}
\multicolumn{5}{p{0.45\textwidth}}{\small {\bf Notes:} \emph{dyn}--rotation curve decomposition, \emph{phot}--SED modelling, \emph{MSC}--marginal stability condition, $\Udisc\!=\!0.5$-- Equation (\ref{eq:Ms_from_MLratio}) with a constant $\Udisc$ of $0.5$. Fits are in the form $\log(y)\!=\!a\,\log(x)+b$, with $x$ and $y$ defined in the Table headers. $\sigma_{M_\star}$ (or $\sigma_{j_\star}$) and $\sigma_\perp$ are the observed vertical and orthogonal scatter, respectively.}
\end{tabular}
\end{table}

Panel ($a$) of Fig.\,\ref{f:phot_vs_dyn} shows the comparison between $M_{\rm phot}$ and $M_{\rm dyn}$ for the $95$ SPARC galaxies in common to both methods.
Clearly, the two $M_\star$ estimates agree remarkably well across almost $5$ orders of magnitude in $M_\star$, with the very few outliers being the galaxies with the poorest photometric fits, quantified by $p(>\!\chi^2)$.
Excluding the eight galaxies with $p(>\!\chi^2)$ below $0.05$ leads to a median $\log(M_{\rm phot}/M_{\rm dyn})$ of $-0.05$ dex with a standard deviation of $0.22$ dex, corresponding to a systematic agreement at the $\sim12\%$ level and $\sim55\%$ scatter across almost $5$ dex in $M_\star$.
The scatter between $M_{\rm phot}$ (or $M_{\rm dyn}$) and the masses derived with the constant $\Udisc$ method is similar, whereas it increases to $\sim0.5$ dex when the MSC method is used (Appendix \ref{a:supplementary} and Fig.\,\ref{f:mass_comparison}).
Excluding systems with a bulge fraction $>0.2$ (purple circles in Fig.\,\ref{f:phot_vs_dyn}) does not change these numbers significantly.
The small systematic offset between $M_{\rm phot}$ and $M_{\rm dyn}$ can be mitigated by lowering $\Udyn$, for example with adiabatic contraction of the DM halos \citep{Li+22}, or by raising $\Uphot$, for example with fewer or fainter AGB stars \citep{Schombert+19} or with a steeper (bottom-heavy) IMF.

Panel ($b$) of Fig.\,\ref{f:phot_vs_dyn} shows the relation between $\Uphot$ (or $\Udyn$) and $L_{3.6}$. 
$\Uphot$ systematically increases with $L_{3.6}$ as a consequence of the typically different SFHs of low- and high-$L_{3.6}$ galaxies; the most luminous galaxies have assembled their mass earlier, and thus are characterised by an overall older (i.e. redder) stellar population and by a higher $\Uphot$ than the low-luminosity ones \citep[the so-called `downsizing', e.g.][]{Cowie+96}.
A similar trend emerges from models with SFHs anchored to the main sequence of star formation for the population of galaxies as a whole \citep{Schombert+19,Schombert+22}. Here, in contrast, the correlation between $\Uphot$ and $L_{3.6}$ emerges from SED modelling of individual galaxies without imposing any external, population-based constraints.
The trend is gentle, as $\Uphot$ varies in the range $0.2$--$0.7$ over almost $4$ dex in $L_{3.6}$, and is conserved if we consider only galaxies without bulges.
Instead, $\Udyn$ does not show a well-defined trend with $L_{3.6}$, possibly because of its larger uncertainties, which range from $\pm0.3$ for low-$L_{3.6}$ galaxies to $\pm0.1$ for the brightest ones.
Excluding galaxies with $p(>\!\chi^2)\!<\!0.05$ leads to a quite narrow distribution of $\Uphot$ (orange histogram in panel ($c$) of Fig.\,\ref{f:phot_vs_dyn}), with a mean (median) of $0.45$ ($0.47$) and a standard deviation of $0.14$. 
For $\Udyn$, these values become $0.54$ ($0.52$) and $0.22$.
Thus both approaches return typical mass-to-light ratios near $0.5$, in agreement with \citet{McGaughSchombert14}.
We have verified that similar trends, but with larger scatter, are visible when using the 2MASS $K_{\rm s}$ band instead of IRAC $3.6\um$ band, leading to a mean (median) $M_{\rm phot}/L_{\rm Ks}$ of $0.71$ ($0.72$) and standard deviation of $0.16$.

We now investigate how different estimates of $M_\star$ affect the scaling laws. 
For simplicity, we focus on the `stellar' TFR and FR, rather than their `baryonic' counterparts, which include both stars and ISM.
For the TFR, we adopt values for $v_{\rm flat}$ from \citetalias{Lelli+16c}, while for the FR we use the specific stellar angular momentum ($j_\star$) estimates from \citet{Posti+18}.
We produced four versions of each relation, using the four methods of estimating $M_\star$ outlined in Session \ref{s:mstar}.
We limited our sample to galaxies that have all four $M_\star$ estimates available, and with $p(>\!\chi^2)>0.05$ in their photometric fit, which leaves us with $68$ ($55$) systems to study the TFR (FR).
This restriction leads to robust comparisons (especially the scatter) but with a reduced sample size.
We have verified that using all galaxies available for each method (which increases the sample size by up to $30$--$80\%$) leads to similar slopes, but a systematically larger scatter.

The resulting scaling relations are shown in Fig.\,\ref{f:scaling_relations}.
We fitted all relations with a power law $y\!=\!x^a 10^b$ hence $\log(y)\!=\!a\,\log(x)+b$, where ($x$,$y$) are ($v_{\rm flat}/150\kms$, $M_\star/\msun$) for the TFR and ($M_\star/10^{10.5}\msun$, $j_\star/\kmskpc$) for the FR, using the \textsc{BayesLineFit} software package in \textsc{python} \citep{Lelli+19}, which provides an MCMC-based Bayesian fit to data with errors in both co-ordinates and finite intrinsic scatter.
The best-fit parameters, together with the observed scatter in the vertical ($\sigma_{M_\star}$ or $\sigma_{j_\star}$) and the orthogonal ($\sigma_\perp$) directions, are listed in Table \ref{t:scaling_relations_params} and in the top left part of each panel of Fig.\,\ref{f:scaling_relations}.

The slopes, $a$, and zero-points, $b$, of the scaling relations agree within twice the quoted uncertainty for all methods except the MSC, which shows some tensions with the others.
We find typical slopes of $a\approx5$ for the TFR (excluding the MSC method) and $a\approx0.6$ for the FR, in line with previous studies \citep[e.g.][]{Lelli+19,FallRomanowsky18}.
The photometric method provides the steepest (shallowest) slope in the TFR (FR) due to $\Uphot$ gradually increasing with $M_\star$, whereas the MSC method shows the opposite behaviour.
The MSC also leads to the largest $\sigma_\perp$ for both scaling relations, which may be partially due to our simplified treatment of the gas component (see Section \ref{ss:other_methods}).
Remarkably, the other methods lead to very similar $\sigma_\perp$, despite featuring different mass-to-light ratios.
Table \ref{t:scaling_relations_params} also shows that, contrary to a suggestion by \citet{Lelli+16a}, the choice of a constant $\Udisc$ is not the only one that minimises the scatter in the TFR.
We have verified that removing the contribution of bulges from our $M_\star$ and $j_\star$ estimates leads to $a$ and $b$ values that are compatible with those of Table \ref{t:scaling_relations_params} within twice the quoted uncertainty.
Combining SPARC with the sample of super-massive ($M_\star>10^{11}\msun$) spirals with extended \hi\ rotation curves from \citet{DiTeodoro+23} leads to fully consistent scaling relations.

\section{Conclusions}\label{s:conclusion}
$M_\star$ estimates that are robust and homogeneous over a broad redshift range are crucial to deriving accurate galaxy scaling relations and their evolution with redshift.
SED modelling techniques using broad-band photometry are routinely employed to infer photometric masses ($M_{\rm phot}$) in the high-$z$ Universe, where the quality of the spectroscopic data is inadequate to determine masses via dynamical methods ($M_{\rm dyn}$). 
However, the two approaches can be compared for galaxies in the local Universe, where both deep, multi-wavelength photometry and high-quality extended \hi\ rotation curves are available.

In this Letter, we have compared $M_{\rm dyn}$ and $M_{\rm phot}$ in a sample of $\sim100$ nearby, star-forming galaxies with accurate \hi\ and H$\alpha$ rotation curves from the SPARC database \citepalias{Lelli+16c}.
Our $M_{\rm dyn}$ estimates were taken from \citetalias{Posti+19a} and were derived using rotation curve decomposition techniques.
We have determined $M_{\rm phot}$ by applying the SED modelling software \bagpipes\ \citep{Carnall+18} to our new photometry using up to ten bands, ranging from the far-ultraviolet to the mid-infrared.

Our results can be summarised as follows.
\begin{itemize}
    \item $M_{\rm dyn}$ and $M_{\rm phot}$ show a median offset of $\sim0.05$ dex and a scatter of $\sim0.22$ dex over almost $5$ dex in $M_\star$ (panel $a$ in Fig.\,\ref{f:phot_vs_dyn}), indicating that the two methods provide systematically compatible estimates.  
    \item Our $M_{\rm phot}$ estimates correspond to mass-to-light ratios in the $3.6\um$ band ($\Uphot$) that increase slowly with the $3.6\um$ luminosity $L_{3.6}$ (panel $b$ in Fig.\,\ref{f:phot_vs_dyn}), an indication of downsizing in the late-type galaxy population.
    \item The choice of using either $M_{\rm phot}$ or $M_{\rm dyn}$ has only a marginal impact on the slope and zero-point of the Tully-Fisher and Fall relations  (Fig.\,\ref{f:scaling_relations} and Table \ref{t:scaling_relations_params}). In particular, the observed orthogonal scatter in both scaling relations is virtually the same for the two methods, and identical to that derived using a constant $\Udisc=0.5$.
\end{itemize}

Our findings have key implications for studies of the evolution of scaling relations, as they validate the use of SED fitting as a powerful tool to infer $M_\star$ at all $z$.
Such techniques can be readily applied to photometry of deep, multi-wavelength images at various redshifts with a uniform set of assumptions in the SED modelling.

\section{Data availability}
Tables C.1 and C.2 are only available in electronic form at the CDS via anonymous ftp to \url{https://cdsarc.cds.unistra.fr/} (130.79.128.5) or via \url{https://cdsarc.cds.unistra.fr/viz-bin/cat/J/A+A/695/L23}.

\begin{acknowledgements}
We thank F.\,Lelli, A.\,Ponomareva, C.\,Spiniello and G.\,van de Ven for insightful discussions, and A.\,C.\,Carnall for providing a version of \bagpipes\ that includes stellar remnants in stellar mass estimates.
AM acknowledges funding from the INAF Mini Grant 2023 program `The quest for gas accretion: modelling the dynamics of extra-planar gas in nearby galaxies'.
EDT was supported by the European Research Council (ERC) under grant agreement no. 10104075.
PEMP acknowledges the support from the Dutch Research Council (NWO) through the Veni grant VI.Veni.222.364
\end{acknowledgements}

\bibliographystyle{aa} 
\bibliography{stellarmass_letter} 

\begin{appendix}
\section{Photometry} \label{a:procedure}
\subsection{Flux measurements} \label{a:photometry}
\begin{sidewaysfigure*}
\begin{center}
\includegraphics[width=1.0\textwidth, trim={4.0cm 10.0cm 4.0cm 0.0cm},clip]{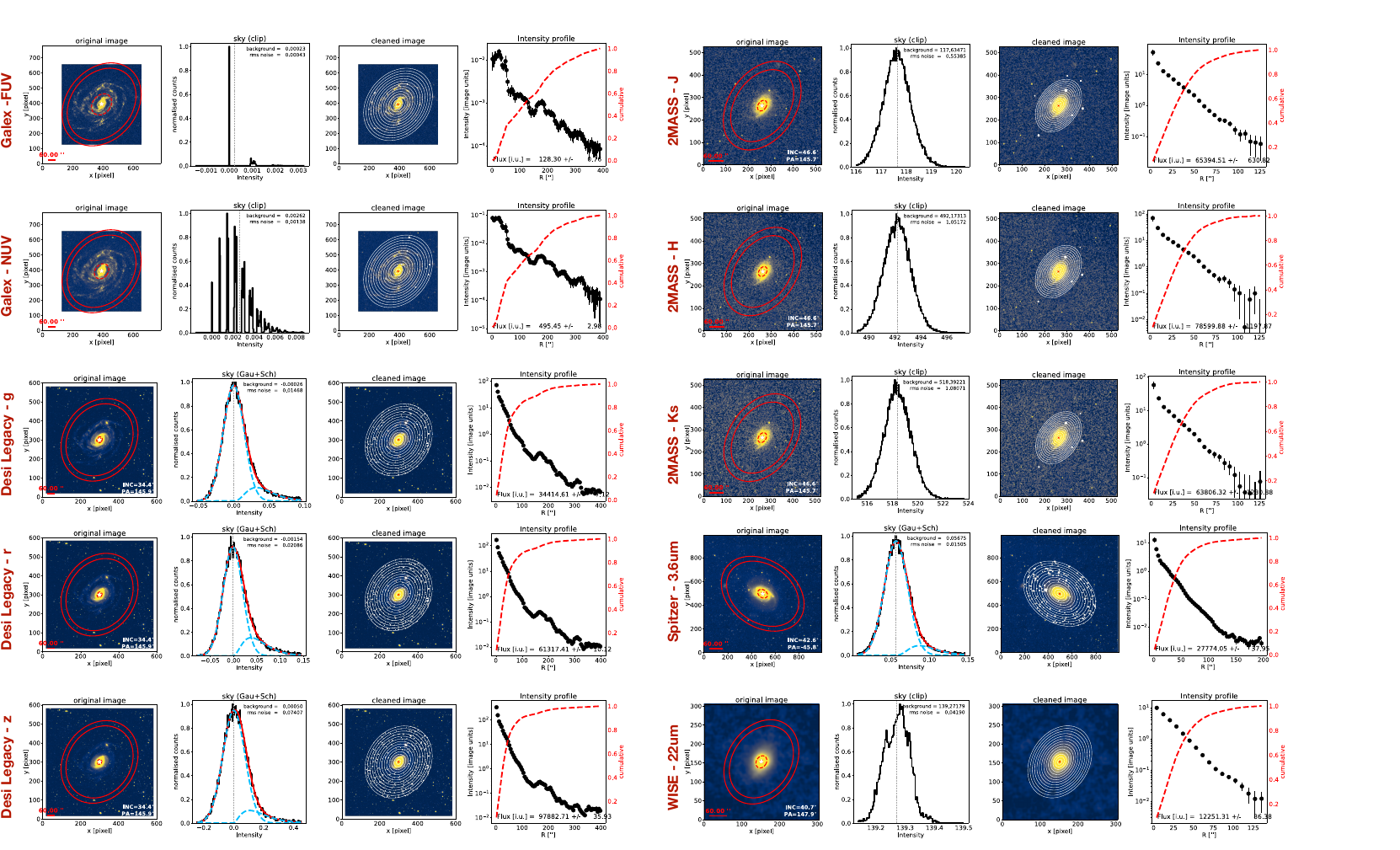}
\caption{Photometric analysis of NGC\,289. Results for each waveband are displayed in two pairs of adjacent panels. \emph{First panel}: original image. The two red ellipses enclose the region where the sky properties are computed. \emph{Second panel}: pixel intensity distribution within the sky region (black histogram). The vertical dotted line indicates the estimated sky background value. The red curve (present only in some bands) shows the best-fit sky model made by the sum of a Gaussian and a Schechter component (see \citetalias{Marasco+23a}), individually shown by dashed light blue curves. \emph{UV} bands suffer from low photon counts. \emph{Third panel}: image filtered with our point-source masking technique. The division in concentric rings is also shown. \emph{Fourth panel:} final radial intensity profile in image units (black circles with error bars) and normalised growth curve (red dashed curve).}
\label{fig:photo_example}
\end{center}
\end{sidewaysfigure*}

\begin{figure}
\begin{center}
\includegraphics[width=0.45\textwidth]{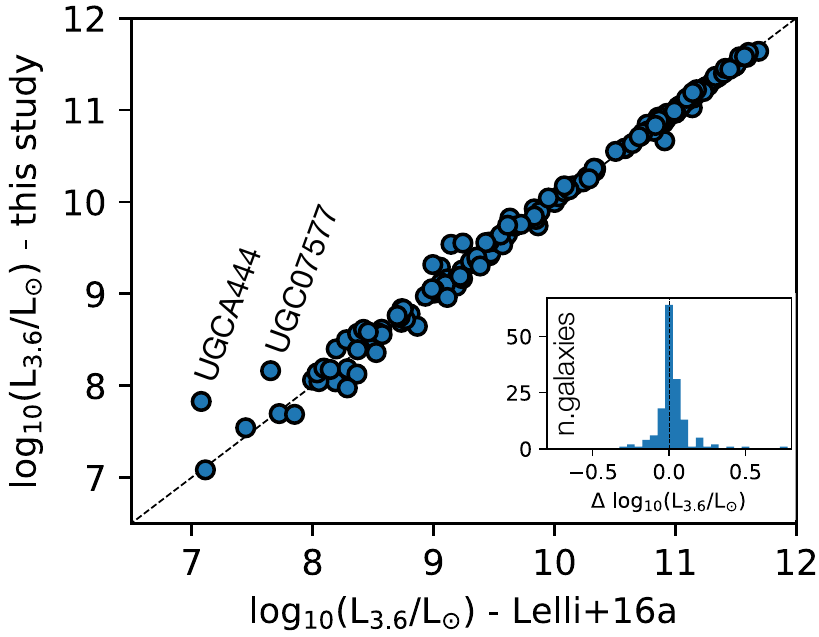}
\caption{Comparison between the $3.6\um$ luminosities of the SPARC galaxies derived in this study and those provided by \citetalias{Lelli+16c}. The inset shows the distribution of the logarithmic difference between the new and the old measurements.}
\label{fig:L36_comparison}
\end{center}
\end{figure}
Our photometry follows the procedure described in Appendix A of \citet{Marasco+23a} (hereafter \citetalias{Marasco+23a}). 
Specifically, we extract the cumulative light profiles from sky-subtracted images after removing contamination by point-like sources such as foreground stars and background galaxies.
Contamination by point-like sources can be particularly severe in IRAC $3.6\um$ images, leading to diverging flux growth curves if not properly accounted for \citepalias[see Figure A2 in][]{Marasco+23a}.
With respect to the original approach of \citetalias{Marasco+23a}, the main novelties introduced are the automatic computation of the galaxy geometrical parameters, the use of a `local' measurement for the sky properties, and a different method to compute the photometric errors. 

The galaxy centre is determined using the IRAC $3.6\um$ image (or, for the systems that are faint in the $3.6\um$ band, using the DESI Legacy $g$-band image) as the final intensity-weighted centre computed in a series of circular apertures of progressively smaller radii.
We start from an initial aperture with radius of $10"$, centred on the co-ordinates provided by the SIMBAD astronomical database \footnote{\url{https://simbad.cds.unistra.fr/simbad/}}, and keep decreasing the radius until the distance between two consecutive centroids is $<0.1$ pixels.
Ideally we would keep the galaxy centre fixed in all bands, but occasionally we need to adjust it manually due to minor imperfections in the astrometric calibration.
The galaxy inclination (INC) and position angle (PA, computed with respect to the image vertical axis) are determined using the same approach as in the SExtractor software \citep{sextractor}, which is based on the image 2nd-order moments\footnote{A detailed description of the SExtractor method can be found at \url{https://sextractor.readthedocs.io/en/latest/Position.html}, see eqs.\,(7)-(17).}.
However, in order to avoid contamination from external sources that could potentially bias the estimate of the geometrical parameters, we compute the 2nd-order moments on a `cleaned' version of the image.
As in the MORPHOT software package \citep{Fasano+12}, this is constructed by comparing pixel-by-pixel the original image ($I_0$) with that rotated by $180\de$ around the galaxy centre ($I_{180}$): if, at a given pixel position, the residual $(I_0-I_{180})$ is larger than $3\times$ the rms intensity determined for $I_{180}$ around the same position, the pixel value in $I_0$ is replaced with the corresponding one in $I_{180}$.
This approach significantly improves the determination of the galaxy shape by removing sources that are asymmetric with respect to the centre.
We determine a unique value for INC and PA for all bands associated with the same instrument (all the optical bands from DESI Legacy, all the near-IR bands from 2MASS, and so on), but allow for variations between different facilities to account for changes in resolution, sensitivity, and overall image quality between one instrument and another.
Variations in INC and PA among the different bands are typically $<10\de$, but can increase in the $22\um$ band due to its poor angular resolution ($12"$).
Finally, we use the galaxy isophotal limit ($R_{\rm max}$) to separate the `sky' region from the `galaxy' region \citepalias[see][]{Marasco+23a}.
Following SExtractor, $R_{\rm max}$ is based on the centred 2nd-order moments of the image, but with respect to the original procedure we multiply it by a factor $2$ (or more, for particularly extended systems) to ensure a more conservative separation between sky and galaxy regions.
As in \citetalias{Marasco+23a}, $R_{\rm max}$ sets the upper limit for the outermost radius in the surface brightness profile. 
However, as we stop tracking the profile when the signal-to-noise ratio within a given ring falls below unity, the actual outermost radius is typically smaller than $R_{\rm max}$.

In \citetalias{Marasco+23a} the sky properties were computed using the entire portion of the image outside the galaxy region.
This approach is ideal for images with a uniform sky, but cannot be applied when sky variations across the field are substantial. 
Here, we prefer to adopt a local approach, where the sky properties are determined in a ring external to the outermost ellipse considered (defined by $R_{\rm max}$).
The size of the annulus is chosen so that it encloses either $20\%$ of the area inside the outermost ellipse or $10^3$ pixels, whichever is larger.

Fig.\,\ref{fig:photo_example} shows the typical outputs of our photometric analysis on the SPARC galaxy NGC\,289.
The flattening reached by the cumulative light profiles in the various bands, shown by the dashed red line in the rightmost panel of each panel-set, is a good indication of the overall quality of our photometric analysis.
In Fig.\,\ref{fig:L36_comparison} we compare the $3.6\um$ luminosities determined from our photometry with the original values provided by \citetalias{Lelli+16c}.
Overall, there is excellent agreement between the two: the median difference between the old and the new values is $0.01$ dex and the standard deviation is $0.12$ dex.
UGC07577 and UGCA444 are the only galaxies for which the two $L_{3.6}$ measurements are in disagreement by more than $0.5$ dex. 
A visual inspection of their IRAC images indicates that these are quite faint but very extended galaxies; thus it is possible that the original photometry did not capture the full extent of their $3.6\um$ light.

\subsection{Flux uncertainties}\label{a:uncertainties}
As in \citetalias{Marasco+23a}, the uncertainty on the total flux ($\epsilon_{\rm flux}$) is given by the quadratic sum of two errors, the first due to the image noise ($\epsilon_{\sigma}$) and the second due to our methodology ($\epsilon_{\rm met}$).
We compute $\epsilon_{\sigma}$ under the assumption of purely Gaussian noise and spatially correlated pixels due to the finite sampling of the point spread function (PSF).
Under these assumptions, we find 
\begin{equation}\label{eq:flux_error}
    \epsilon_{\sigma} = \sigma_{\rm sky} (2\,n_{\rm pix} n_{\rm res})^{1/2}, 
\end{equation}
where $\sigma_{\rm sky}$ is the sky noise, $n_{\rm pix}$ is the number of pixels inside the outermost ellipse considered, and $n_{\rm res}$ is the number of pixels within a single resolution element. 
The latter is computed assuming a 2D Gaussian shape for the PSF with known full width at half maximum (FWHM in pixel units):
\begin{equation}\label{eq:n_res}
    n_{\rm res} = {\rm max} \left[\frac{\pi}{4\ln(2)}({\rm FWHM})^2,\frac{1}{2}\right]\,.
\end{equation}
We have tested Equation (\ref{eq:flux_error}) using mock images, finding that it correctly describes the uncertainty on the flux for FWHM $>1$ pixel.
We note that, for an undersampled PSF, $n_{\rm res}\simeq1/2$ thus $\epsilon_\sigma\propto\sqrt{n_{\rm pix}}$, as expected from Gaussian noise added to independent elements.
However, the assumption of Gaussian noise may not be valid for the photon-limited \emph{GALEX} images (see, for instance, the sky intensity distribution in the \emph{FUV} and \emph{NUV} bands in Fig.\,\ref{fig:photo_example}), in which case $\epsilon_{\sigma}$ may be underestimated. 
Flux measurements below $3\epsilon_\sigma$ are considered non-detections: upper-limits are computed as $3\times$ the $\epsilon_\sigma$ from Equation (\ref{eq:flux_error}) for $n_{\rm pix}\!=\!1$.

In \citetalias{Marasco+23a}, $\epsilon_{\rm met}$ was determined for each galaxy as the standard deviation of the flux distribution derived by randomising the various parameters of the photometric method in given ranges.
Fluctuations in the values of these parameters, which regulate the geometry of the elliptical aperture, the estimate of the sky properties, and the size of the mask, can affect flux measurements more than the image noise alone, especially in the high signal-to-noise regime.
We have applied the same technique to a few galaxies in the SPARC sample, finding a variable $\epsilon_{\rm met}$ ranging from $\sim1\%$ to $\sim10\%$ of the measured flux, depending on the properties of the image itself.
In line with these findings, and for simplicity, here we have fixed $\epsilon_{\rm met}$ to a constant $5\%$ of the measured flux for all bands.
This is larger than the mean $\epsilon_{\sigma}$ of the sample ($\sim1.5\%$ of the flux) and is the dominant source of uncertainty in our photometric measurements.
In fact, only $7\%$ of the photometric data used for SED modelling have $\epsilon_{\sigma}\!>\!\epsilon_{\rm met}$.

We correct for foreground extinction in the Milky Way (MW) using the reddening maps of \citet{SchlaflyFinkbeiner11} and the $R_{\rm V}$-dependent extinction curve of \citet{Fitzpatrick99}, assuming $R_{\rm V}\!=\!3.1$.
We have checked the impact of variations in $R_{\rm V}$ on our extinction-corrected fluxes using a Monte-Carlo approach, where we randomly extract $R_{\rm V}$ from a Gaussian distribution centred on the value assumed and with a standard deviation of $0.25$, which approximates the $R_{\rm V}$ spread in the MW \citep{Zhang+23}.
We find typical variations in the UV-band fluxes (which are those maximally affected by the correction) of $\sim1\%$, and always smaller than our photometric uncertainty.

We stress that having a minimum flux error given by a fixed fraction of the total flux helps to avoid over-weighting bands with unrealistically small error bars in the \bagpipes\ fitting, providing a more balanced weighting over the wavelength range explored.
In addition, the fact that the $p(>\!\chi^2)$ values determined after the SED modelling are not systematically close to $0$ (i.e. model incompatible with the data) or to $1$ (i.e. model overfits the data) is an indication that our photometric errors are neither underestimated or overestimated.

\section{SED modelling} \label{a:SED_modelling} 
\begin{figure*}
\begin{center}
\includegraphics[width=0.9\textwidth]{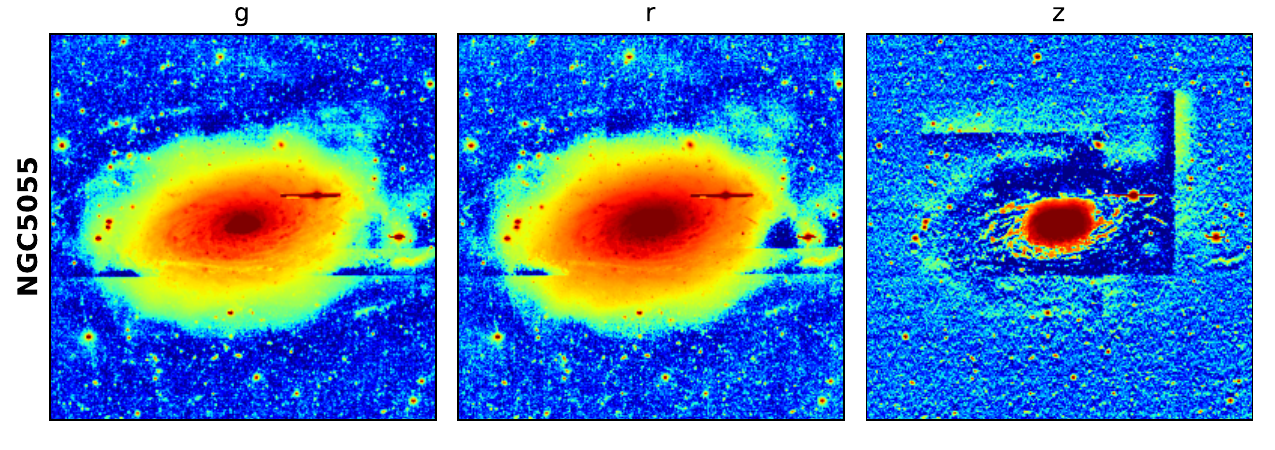}
\includegraphics[width=0.9\textwidth]{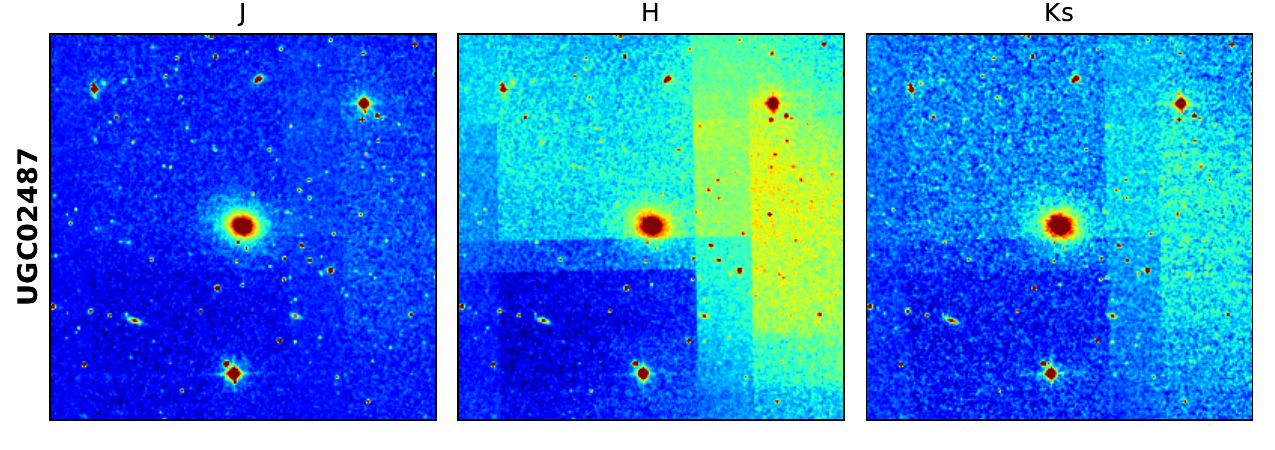}
\caption{Example of images affected by artifacts in our galaxy sample. \emph{Top panels}: DESI Legacy images of NGC\,5055. The $z$ band image suffers from severe calibration problems, causing an underestimate of the flux. \emph{Bottom panels}: NIR images of UGC\,02487 from 2MASS. The $H$ and the $K_{\rm s}$ bands show a variable sky background that can lead to errors in the derived flux.} 
\label{f:badimages}
\end{center}
\end{figure*}

\begin{figure}
\begin{center}
\includegraphics[width=0.4\textwidth]{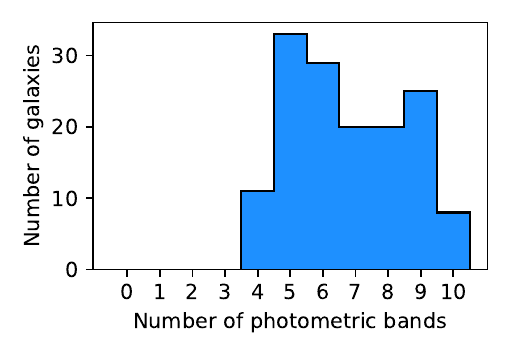}
\caption{Histogram of the number of photometric bands used in our SED modelling of the SPARC sample.} 
\label{f:nbands_photo_histo}
\end{center}
\end{figure}

\begin{figure*}
\begin{center}
\includegraphics[width=0.75\textwidth]{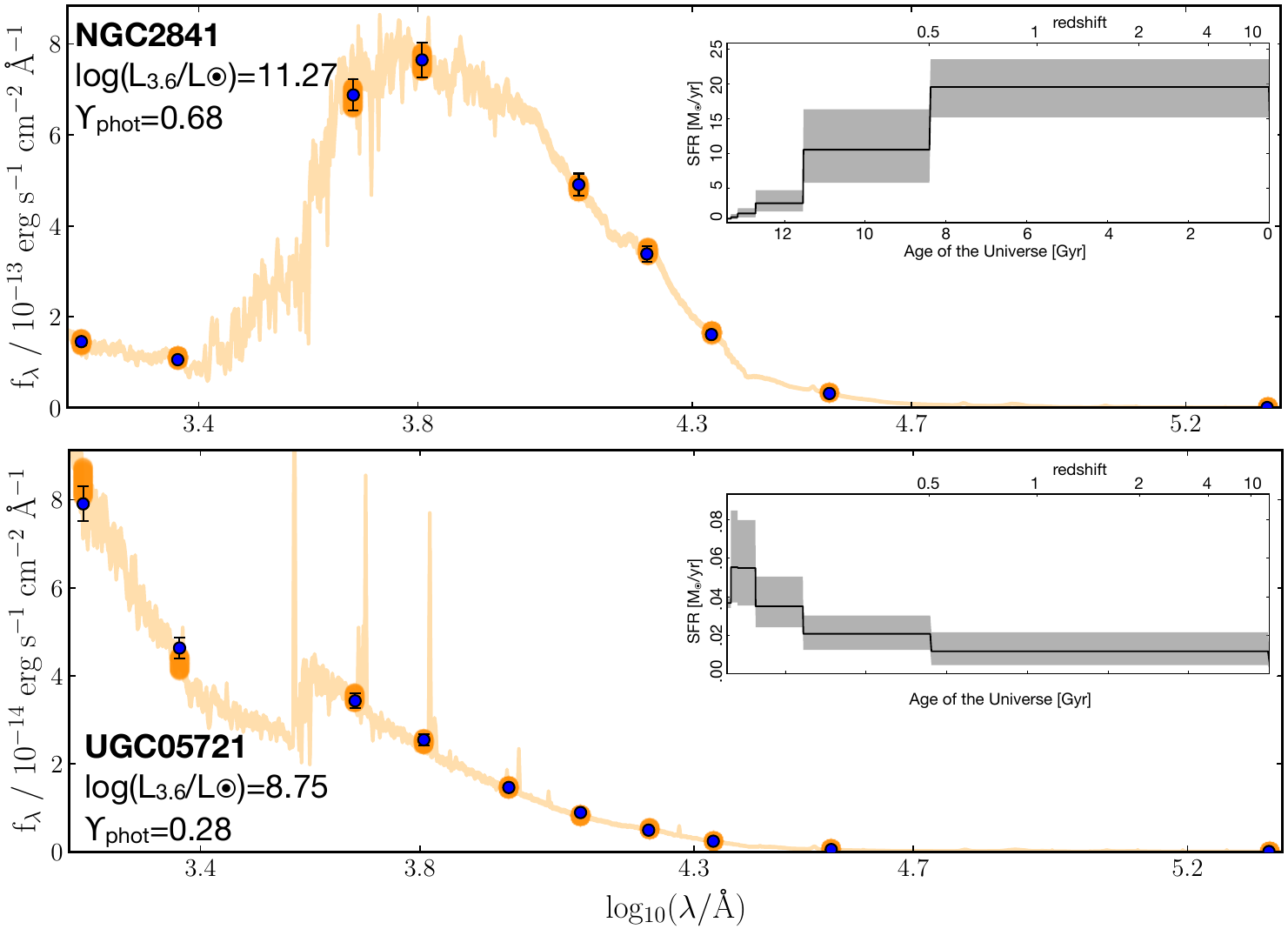}
\caption{Two examples of our SED modelling using \bagpipes: NGC\,2841 (top) and UGC\,05721 (bottom). The blue points with error bars show our measured fluxes in the various bands ($FUV$, $NUV$, $g$, $r$, $z$, $J$, $H$, $K_{\rm s}$, $3.6\um$, $22\um$). The $z$ band is used only in UGC\,05721.
The orange-shaded region encompasses the set of SED models that best reproduce the data, with the dark orange points showing the predicted fluxes in the same wavebands as the measurements.
The insets show the SFH of the model set. The two galaxies have markedly different SFHs and corresponding values of $\Uphot$, listed on the left side of each panel. } 
\label{f:example_bagpipes}
\end{center}
\end{figure*}

\begin{figure*}
\begin{center}
\includegraphics[width=0.8\textwidth]{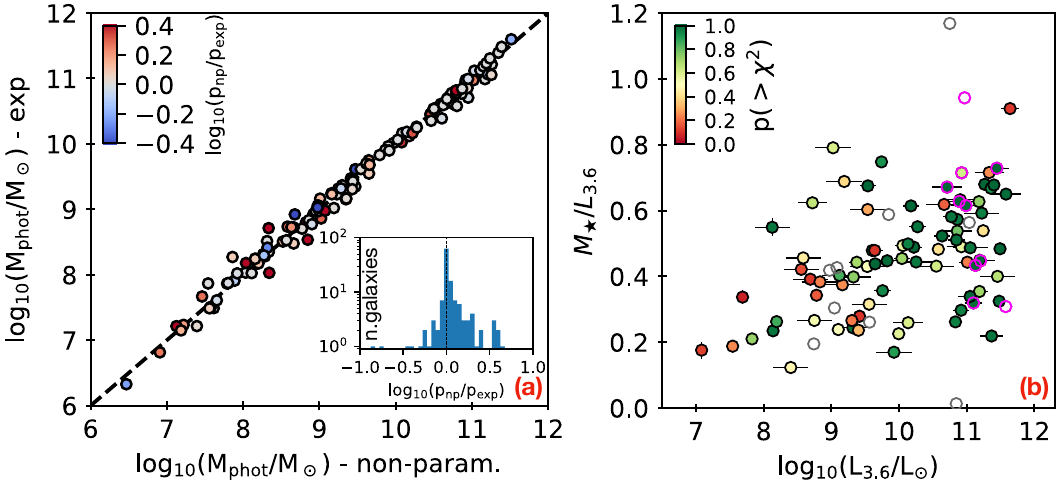}
\caption{($a$)--comparison between $M_{\rm phot}$ for the best-fitting exponential and non-parametric SFHs, colour-coded by the ratio of $p(>\!\chi^2)$ associated with the two models. Red colours indicate higher $p(>\!\chi^2)$, hence higher quality of fit, for the non-parametric SFH. The inset shows the distribution of the $p(>\!\chi^2)$ ratio, clearly skewed towards positive values. ($b$)--correlation of $M_\star/L_{3.6}$ with $L_{3.6}$ for models with exponential SFHs. The trend is similar to the one for non-parametric SFHs shown in Fig.\,\ref{f:phot_vs_dyn} but with larger scatter.} 
\label{f:nonparam_vs_param}
\end{center}
\end{figure*}

The $10$ photometric bands we analyse provide information on the recent star formation ($FUV$, $NUV$ and $22\um$ bands), on the intermediate-age stars (optical bands) and on the old stellar population (near-IR bands), allowing us to infer SFHs via SED modelling.
In our \bagpipes\ modelling we make use of a non-parametric SFH in $7$ age bins: $0$-$30\Myr$, $30$-$100\Myr$, plus additional 5 logarithmically spaced bins between $100\Myr$ and the age of the Universe.
We checked that increasing the number of age bins has virtually no impact on our estimates of $M_{\rm phot}$.
To avoid abrupt discontinuities in the star formation rates between different age bins we make use the `continuity' prior based on Student's t-distribution described in \citet{Leja+19} and implemented in \bagpipes.
This method explicitly weights against abrupt changes in SFR$(t)$, while remaining flexible enough to model a wide range of SFHs.

\bagpipes\ employs the 2016 version of the \citet{BruzualCharlot03} SPS models and the IMF of \citet{KroupaBoily02}.
We adopt the dust attenuation model of \citet{CF00}, using a value of $0.7$ for the power law slope of the attenuation law and a multiplicative factor of $3$ for the $A_{\rm V}$ associated with the birth clouds (which in \bagpipes\ have a lifetime of $10\Myr$).
We tested other attenuation laws finding compatible $M_{\rm phot}$, but overall marginally lower-quality fits.
We treat the $A_{\rm V}$ of the stellar populations older than $10\Myr$ as a free parameter, adopting a uniform prior in the range $0\!<\!A_{\rm V}\!<\!2$.
The model spectra include nebular emission lines based on the \textsc{cloudy} photoionisation code \citep{Ferland+17}, but for simplicity we fix the ionisation parameter to $\log(U)\!=\!-3$, which provides line fluxes compatible with those observed in typical star-forming galaxies \citep[see][]{Carnall+18}.
Given the lack of powerful starburst galaxies in the SPARC sample, our choice of $\log(U)$ has a very small impact on the broad-band model SED.
An additional free parameter used in our fit is the stellar metallicity $Z$, for which we use a unique value for all age bins and assume a uniform prior in the range $0.1\!<\!\log(Z/Z_\odot)\!<\!2$.
The lack of spectroscopic information implies that $Z$ remains poorly constrained by the model, so it is treated here as a `nuisance parameter' over which the posterior distributions of the other (and more relevant) parameters are marginalised.

Model fitting in \bagpipes\ is implemented within the framework of the Bayesian inference, and is based on comparing the (MW extinction-corrected, see Appendix \ref{a:uncertainties}) observed and predicted SEDs.
The latter are obtained by applying the appropriate broad-band filters of the various observational facilities\footnote{available at \url{http://svo2.cab.inta-csic.es/theory/fps/}} to the model spectra. 
Further details on the fitting scheme of \bagpipes\ are provided in Section 3.2 of \citet{Carnall+18}.

As discussed in Section \ref{ss:mstar_photo}, in our SED modelling we use only observations that satisfy a series of quality flags.
First, we require a signal-to-noise ratio on the total flux larger than $5$.
Second, to ensure that we are not missing a significant fraction of the light because of an insufficient sensitivity or an incorrect modelling of the sky properties, we impose a condition on the flattening of the cumulative light profile $I(<R)$ determined with our photometric analysis.
Following \citet{Posti+18}, we require that the last two points of the cumulative $I$ profile differ by less than $10\%$ and that the logarithmic slope of the profile in the outermost point is less than $0.5$:
\begin{equation} \label{eq:criteria}
\frac{I(<R_N) - I(<R_{N-1})}{I(<R_N)} < 0.1 \quad \& \quad
\left.\frac{{\rm d}\log I(<R)}{{\rm d}\log R}\right|_{R_N} < 0.5\,.
\end{equation}
Lastly, we visually inspect all our images and flag cases that show clear artifacts that could potentially impact our photometry.
This step is particularly crucial for the DESI $z$ band and for the 2MASS bands (especially $H$ and $K_{\rm s}$), which often show features attributable to imperfect data reduction.
We provide an example in Fig.\,\ref{f:badimages}, which shows two galaxies with artifacts in their optical and NIR images: in this case we have decided to exclude the $z$ band in NGC\,5055 and both the $H$ and $K_{\rm s}$ bands in UGC\,02487.
Measurements that do not obey any of these quality flags are still listed in Table C.1 but are not used in our SED modelling.
In addition to these quality flags, for SED fitting we require a minimum number of $4$ (unflagged) measurements per galaxy.
The typical number of bands per galaxy used in our \bagpipes\ modelling is $7$, with the full distribution shown in Fig.\,\ref{f:nbands_photo_histo}.

Fig.\,\ref{f:example_bagpipes} show the \bagpipes\ fit for two representative galaxies in our sample, NGC\,2841 (top panel) and UGC\,05721 (bottom panel), with photometry in $9$ and $10$ bands, respectively.
The models reproduce the data remarkably well at all wavelengths. 
These two galaxies have very different SFHs and values of $\Uphot$.
NGC\,2841 features a markedly declining SFR; its light is dominated by an old, faint stellar population, which boosts the $\Uphot$ to $\approx0.66$.
In contrast, UGC\,05721 has an increasing SFR that makes this galaxy exceptionally luminous and blue, with $\Uphot\approx0.28$.
NGC\,2841 and UGC\,05721 exemplify nicely the wide range of SFHs exhibited by nearby disc galaxies, which leads to the downsizing effect mentioned in Section \ref{s:results}.

The strengths of SPS models with non-parametric SFH over those with parametric SFHs have been discussed by \citet{Leja+19}.
As a check, we have also modelled the SPARC photometry with the standard exponentially declining SFR parametrised by the e-folding time-scale and the birth time of the galaxy.
The comparison between the parametric and non-parametric $M_{\rm phot}$ estimates is shown in the left panel of Fig.\,\ref{f:nonparam_vs_param}.
In general, there is excellent agreement between the two $M_{\rm phot}$, but estimates that use exponentially declining SFRs come with lower $p(>\!\chi^2)$ (i.e. fits with lower quality) compared to the non-parametric models. 
Additionally, a comparison of Figs.\,\ref{f:phot_vs_dyn} and \ref{f:nonparam_vs_param} reveals that the scatter in the correlation between $\Uphot$ and $L_{3.6}$ is larger when parametric SFHs are used. 
Thus the non-parametric SFHs appear to be more realistic than the parametric ones.

\section{Comparison of stellar masses}\label{a:supplementary}
\begin{figure*}
\begin{center}
\includegraphics[width=0.95\textwidth]{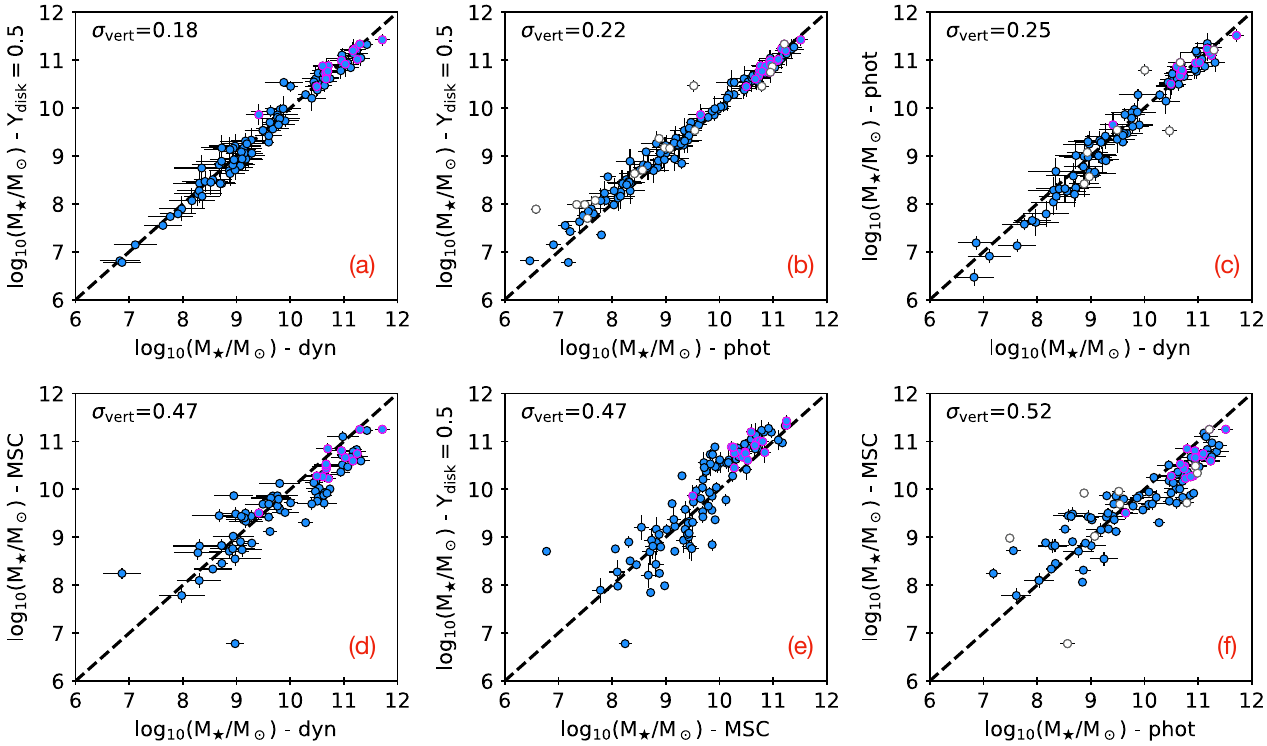}
\caption{Comparison of the different $M_\star$ estimates used in this study. \emph{dyn}--dynamical method based on rotation curve decomposition (Section \ref{ss:mstar_dyn}); \emph{phot}--photometric method based on SED modelling (Section \ref{ss:mstar_photo}); \emph{MSC}--marginal stability condition; $\Udisc\!=\!0.5$--fixed mass-to-light ratio in the $3.6\um$ band. Panels are ordered by increasing vertical observed scatter $\sigma_{\rm vert}$ as indicated in the upper-left corners.}
\label{f:mass_comparison}
\end{center}
\end{figure*}

Fig.\,\ref{f:mass_comparison} compares the $M_\star$ estimates from the four different methods used in this study.
The tightest correlation is the one between the dynamical $M_\star$ and the masses determined using $\Udisc\!=\!0.5$.
This is partially driven by the few galaxies with $M_\star\lesssim10^{8.7}\msun$, for which the $M_{\rm dyn}$ estimate is highly uncertain since the stars are dynamically subdominant relative to the cold gas and the DM.
In these cases, the $\Udisc$ estimated by \citetalias{Posti+19a} scatters around $\simeq0.6$ as this value falls in the middle of the range adopted for the prior ($0.01$--$1.2$).
If we exclude galaxies with $M_\star\!<\!10^{8.7}\msun$ and $p(>\!\chi^2)\!<\!0.05$, we get $\sigma\!=\!0.19$ in panel ($a$) and $\sigma\!=\!0.22$ in panel ($c$), which again corroborates the similarities of the various methods.
We reiterate that the scatter visible using the MSC method can be partially caused by our simplified treatment of the ISM, as mentioned in Section \ref{ss:other_methods}.

\end{appendix}
\end{document}